\documentclass[aps,prb,twocolumn,superscriptaddress,showpacs,amsmath,amssymb,footinbib,longbibliography]{revtex4-1}
\usepackage[english]{babel}
\usepackage{amssymb,amsbsy,amsmath}
\usepackage{graphicx}
\usepackage{dcolumn}
\usepackage{bm}
\usepackage{subfigure}
\usepackage{color}
\usepackage{textcomp}
\usepackage[colorlinks,bookmarks=false,citecolor=blue,linkcolor=red,urlcolor=blue]{hyperref}
\newcommand{\op}[1]{%
    \fontdimen12\textfont3=2pt\fontdimen12\scriptfont3=1.4pt%
    \!\null\mathop{\vphantom{#1}\smash{#1}}\limits_{\sim}\null\!}

\newcommand{\figref}[1]{Fig.~\protect\ref{#1}}

\newcommand*{\opDag}[1]{\op{#1}^{\dagger}}

\newcommand*{\opVec}[1]{\op{\bm{#1}}}

\begin{document}

\title{Magnetization curves of deposited finite spin chains}

\author{Henning-Timm \surname{Langwald}}
\email{langwald@physik.uni-bielefeld.de}
\author{J\"urgen \surname{Schnack}}
\email{jschnack@uni-bielefeld.de}
\affiliation{Fakult\"at f\"ur Physik, Universit\"at Bielefeld, Postfach 100131, D-33501 Bielefeld, Germany}

\date{\today}

\begin{abstract}
The characterization and manipulation of deposited magnetic
clusters or molecules on surfaces
is a prerequisite for their future utilization. In recent years techniques like
spin-flip inelastic electron  
tunneling spectroscopy using a scanning tunneling microscope proved to
be very precise in determining e.g. exchange constants in deposited 
finite spin chains in the meV range. In this article we tackle
the problem numerically by
investigating the transition from where a pure spin Hamiltonian
is sufficient to the point where 
the interaction with the surface significantly alters the magnetic properties.
To this end we study the static, i.e. equilibrium impurity
magnetization of antiferromagnetic chains for varying couplings
to a conduction electron band of a metal substrate. 
We show under which circumstances the screening of a part of the system
enables one to deduce molecular parameters of the remainder from
level crossings in an  
applied field.
\end{abstract}


\maketitle

\section{Introduction}

The question whether and how deposited
magnetic clusters or molecules change their magnetic properties
when deposited on a metallic substrate  
is of fundamental importance especially in view of possible applications 
as next generation storage devices or magnetic logic
circuits.\cite{HFH:S06,HLO:S07,WBL:NM07,THS:JPCM09,MPS:NM09,WYZ:PRL09,BMW:PRL09,BAK:PRL10,MJT:NC16,CRY:NL17,SOX:SA17,Ter:PSS17} 
One experimental method to investigate local magnetic properties
is spin-flip inelastic electron tunneling spectroscopy with a
scanning tunneling microscope.\cite{HFH:S06,CFJ:PRL08} 
In this method, jumps of the differential conductivity signal transitions between 
magnetic levels of the deposited entity, of course under the assumption 
of a weak coupling to the substrate.
When interpreting the observed structure of magnetic levels in such 
measurements one has to conjecture whether and how much of the deposited 
spin system is screened by the conduction electrons, since this has a
great influence on the magnetic properties of the remainder.

\begin{figure}[ht!]
\centering
\includegraphics*[clip,width=55mm]{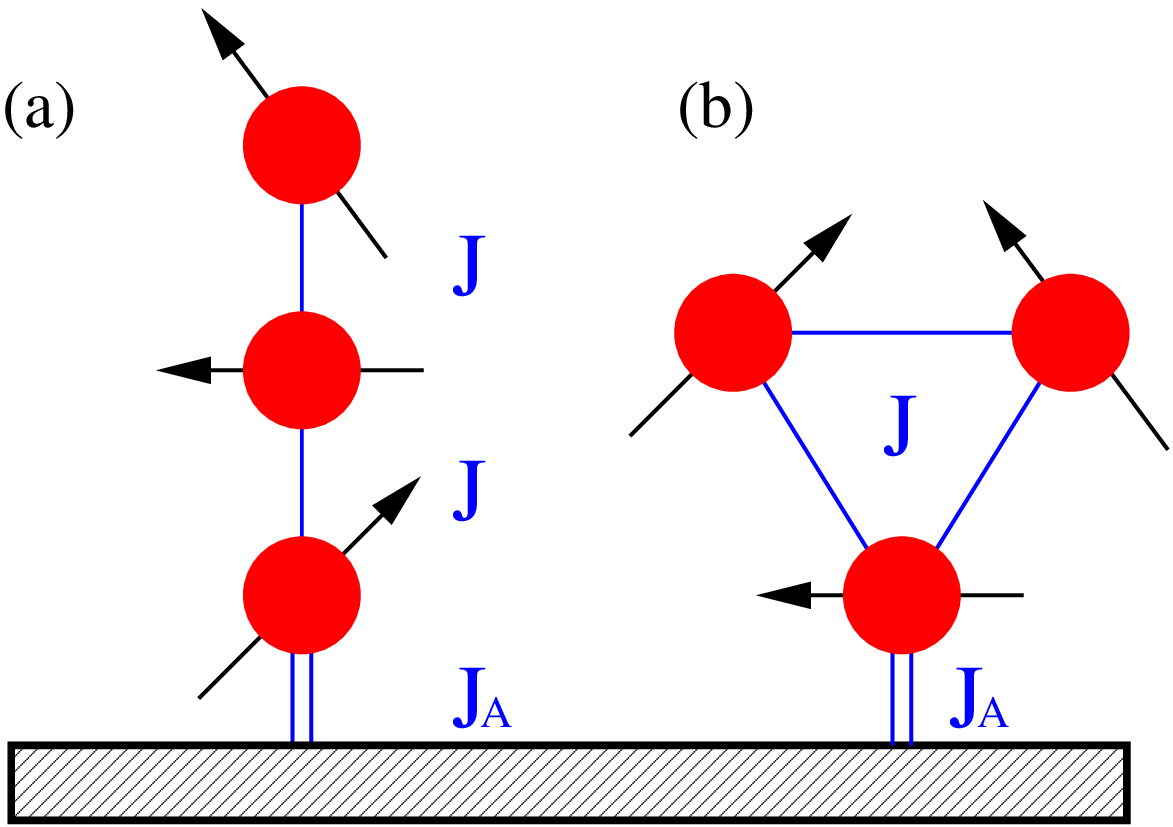}
\caption{(Color online) Sketch of the trimer configuration: stacked magnetic centers interact
  via an exchange coupling $J$, the lowest moment is coupled to
  the metal surface with strength $J_A$. $J$ and $J_A$ are
  unknown, $J$ shall be derived from experimental data. (b)
  shows a triangular configuration discussed later in this
  article.}  
\label{molsurf-f-a}
\end{figure}

In this article we investigate by means of the Numerical Renormalization
Group method (NRG)  numerically
exactly,\cite{BCT:RMP08,Wil:RMP83,Wil:RMP75} how the static,
i.e. equilibrium magnetic properties of deposited spin structures as 
those shown in \figref{molsurf-f-a} depend on the exchange coupling $J_A$ 
to the substrate's conduction electrons, the internal exchange
coupling $J$, and the applied magnetic field $B$ at a fixed very
low temperature $T$.
As one of our model spin structures deposited on the substrate, we want
to take a spin chain of spins $s=1/2$ , an object which
experimentally can be formed by stacking layers of cobalt
phthalocyanine molecules.\cite{CFJ:PRL08}
This arrangement is sketched in \figref{molsurf-f-a}~(a). 

For our investigations we employ a single-channel
single-impurity Kondo model  
as a minimal model to describe 
the correlations induced by the exchange interaction between the
conduction electrons of the non-magnetic metal and one spin of the 
magnetic molecule.\cite{RWH:PRL06A,RWH:PRL06B} As already
demonstrated in Ref.~\onlinecite{HoS:PRB13} the magnetization
steps due to ground state level crossings 
in an applied magnetic field can be used as fingerprints of the 
underlying spin model. This will be shown by   
investigating different sizes of chains as well as triangular
arrangements of magnetic moments, see \figref{molsurf-f-a}~(b).

We would like to stress that our investigations focus on
the equilibrium isothermal impurity magnetization as a function
of the applied magnetic field.
Other studies investigated properties as a function of temperature,
partly within the Kondo model \cite{Zhi:JPCM10}
or more often in the closely related Anderson model.\cite{MJG:JPCB13,MJL:PRB09,Zhi:PRB10,WGL:PRB11,HGA:PRB15}
There the focus is often on the non-equilibrium, i.e. transport
properties of similar structures such as quantum dots,\cite{AnS:PRL05,And:PRL08,NgC:PRB14}
or more recently on impurities coupled to superconducting
leads, see \onlinecite{GSJ:NC18} for a very recent example.
The structure shown in \figref{molsurf-f-a}~(b), for example, 
was investigated with the different focus
on the influence of transport on entanglement in a field-free
arrangement in Ref.~\onlinecite{TBZ:EPJB14}.

As a one-line summary of our investigations we can state, that
under typical experimental situations for typical magnetic
molecules as sketched in \figref{molsurf-f-a} the full screening
of the lowest spin interacting with the substrate requires an
antiferromagnetic exchange coupling larger than
0.2~eV. Otherwise partial screening occurs. 
A very recent experimental investigation, where a singly
occupied molecular orbital was considered, showed that this
value of 0.2~eV is within the potential magnitude of a coupling
to the substrate.\cite{ZSK:ARXIV18} The experiment featured a 
variation of the coupling strength (in the 
form of the Kondo temperature) in similar fashion to the
theoretical discussion here. 

The article starts with a short reminder of NRG in
Sec.~\ref{sec-2}, presents our results in Sec.~\ref{sec-3}  and
in  Sec.~\ref{sec-Ana}, before 
it summarizes our main points in  Sec.~\ref{sec-4}.

\section{Reminder on NRG}
\label{sec-2}

In order to model a molecule and its
coupling to a  
surface we use the following Hamiltonian which consists 
of three parts:\cite{RWH:PRL06A,RWH:PRL06B,HoS:PRB13}
\begin{eqnarray}
\label{E-2-2}
\op{H} &=& \op{H}_{\text{electrons}} + \op{H}_{\text{coupling}} + \op{H}_{\text{impurity}}  \ ,
\\
\label{E-2-3}
\op{H}_{\text{electrons}} &=& \sum_{i \neq j, \, \sigma}{t_{ij}
  \opDag{d}_{i\sigma} \op{d}_{j\sigma}} + g \mu_B B \sum_{i}\;
\op{s}_i^z  \; . 
\end{eqnarray}
The first part $\op{H}_{\text{electrons}}$ represents non-interacting electrons on a lattice. 
The hopping parameter $t_{ij}$ is non-zero only if the lattice sites $i$, $j$ are nearest neighbors. 
$\opDag{d}_{i\sigma}$ and $\op{d}_{j\sigma}$ are fermionic creation 
and annihilation operators for electrons with spin direction $\sigma$. 
The interaction with an external magnetic field $B$ is given 
by the Zeeman term with $\op{s}_i^z $ representing 
the effective electron spin at lattice site $i$, $g$ the g-factor and  
$\mu_B$ the Bohr magneton.
The second part $\op{H}_{\text{impurity}}$ models the impurity, 
i.e., the molecule or chain via an effective Heisenberg model 
for all connected spins $\opVec{S}_i$ and a Zeeman term. 
$J_{ij}$ is the interaction between spins $i$ and $j$
and antiferromagnetic for $J_{ij}>0$, 
\begin{eqnarray}
\label{E-2-1}
\op{H}_{\text{impurity}}
&=&
2
\sum_{i<j}\; J_{ij}
\opVec{S}_i \cdot \opVec{S}_j
+
g\, \mu_B\, B\,
\sum_{i}\;
\op{S}^z_i
\ .
\end{eqnarray}
The last part  $\op{H}_{\text{coupling}}$  describes the 
Kondo-like interaction of the molecule with the surface. 
The coupling constant $J_A$ is positive for antiferromagnetic coupling
\begin{equation}
\label{E-2-4}
\op{H}_{\text{coupling}} = 2 \cdot J_A \cdot \opVec{S}_1 \cdot \opVec{s}_0 \; .
\end{equation}
To calculate thermodynamic values within this model we use 
Wilson's Numerical Renormalization Group (NRG) with 
the discretization scheme proposed 
by \v{Z}itko and Pruschke \cite{ZhP:PRB09,Zhi:CPC09}
and a $z$-averaging for 2 values. A constant density of states
is assumed.

It is possible that in reality molecular 
orbitals like those of phthalocyanine molecules hybridize with surface
states, compare 
e.g. Refs.~\onlinecite{SBM:JACS01,BSH:L04,ZLC:S05,GJH:PRL07,BAK:PRL10,SMM:PRB11}.
Our approach, like others,\cite{RWH:PRL06A,RWH:PRL06B} simplifies the situation to 
a point where the deposited molecule is reduced to its spin
degrees of freedom which interact with the metal's conduction
electrons.

\section{Results and interpretation}
\label{sec-3}

\begin{figure}[ht!]
\centering
\includegraphics*[clip,width=40mm]{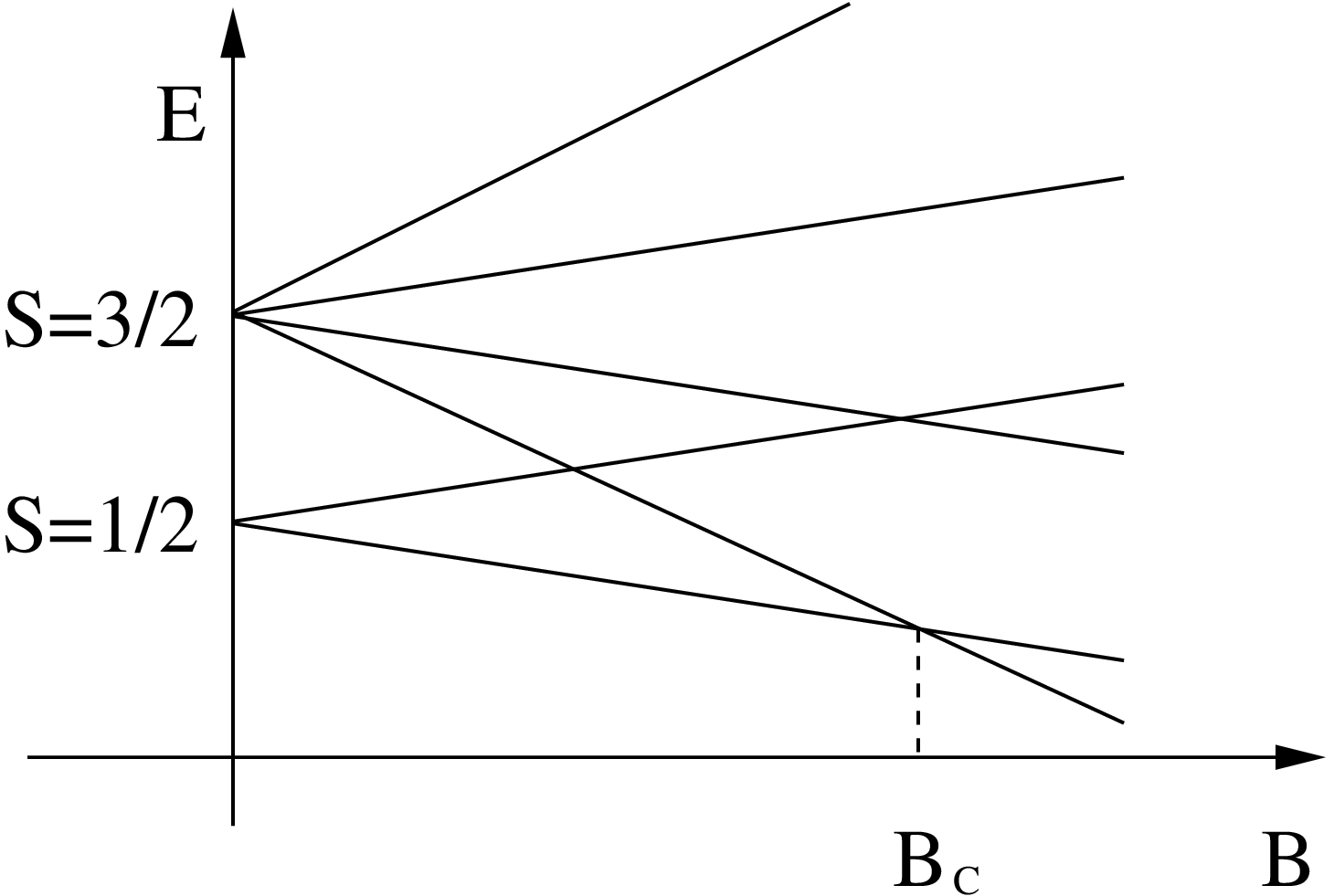}
\includegraphics*[clip,width=40mm]{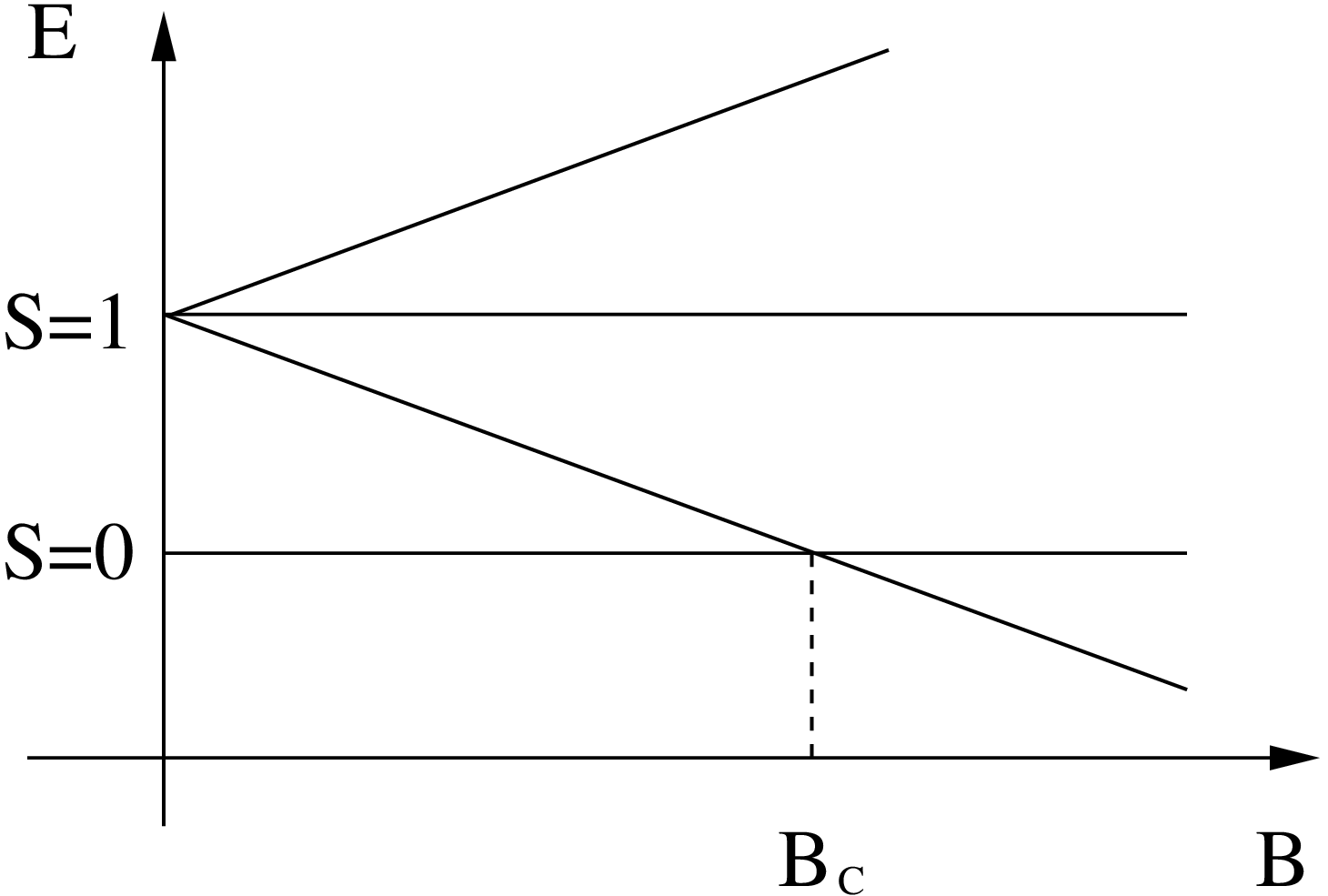}
\caption{(Color online) Sketch of the Zeeman levels of a
  trimer of three spins $s=1/2$ (l.h.s.) and of a dimer of two
  spins $s=1/2$ (r.h.s.).} 
\label{molsurf-f-c}
\end{figure}

Before we discuss our
numerical results of a finite spin chain interacting with a
metal substrate we would like to shortly look at a free
three-site chain of spins $s=1/2$. Its levels split under the
influence of an applied magnetic field as depicted on the
l.h.s. of \figref{molsurf-f-c}. At a certain magnetic field
value $B_c=3\cdot J/(g \mu_B)$ the lowest $(S=1/2,M=-1/2)$ and
$(S=3/2,M=-3/2)$ 
levels cross. If the lowest spin of the trimer would be
completely screened, 
the remaining dimer would possess a different level scheme with
a different crossing field $B_c=2\cdot J/(g \mu_B)$ as depicted
on the r.h.s. of \figref{molsurf-f-c}. Therefore, inferring 
$J$ from spectroscopic data or equivalently from crossing fields
strongly depends on the degree of screening. 

In the following we discuss the so-called impurity magnetization,
i.e. the thermal expectation value of molecular magnetic moment
$\sum_{i}\;\op{S}^z_i$ as a function of the applied field for
various couplings to the substrate. In order to work with
reasonable numbers we set the intramolecular coupling to
$J=1$~meV and the half-bandwidth of the metal to $W=1$~eV. As
the temperature we choose $T\approx 2 \cdot 10^{-4}\ W/k_B
\approx 2.36$~K, which is lower than the intramolecular
coupling, but not too low so that thermodynamic functions are
still smooth.\footnote{Since the thermodynamic functions have to
  be evaluated for every magnetic field value, too abruptly
  varying functions would need more supporting points. Already
  now each NRG curve in any of the figures needs a rather long
  time on a supercomputer.}
In addition, this temperature reflects the relevant experimental
scale of the related STM investigations.\cite{CFJ:PRL08}

\begin{figure}[ht!]
\centering
\includegraphics*[clip,width=85mm]{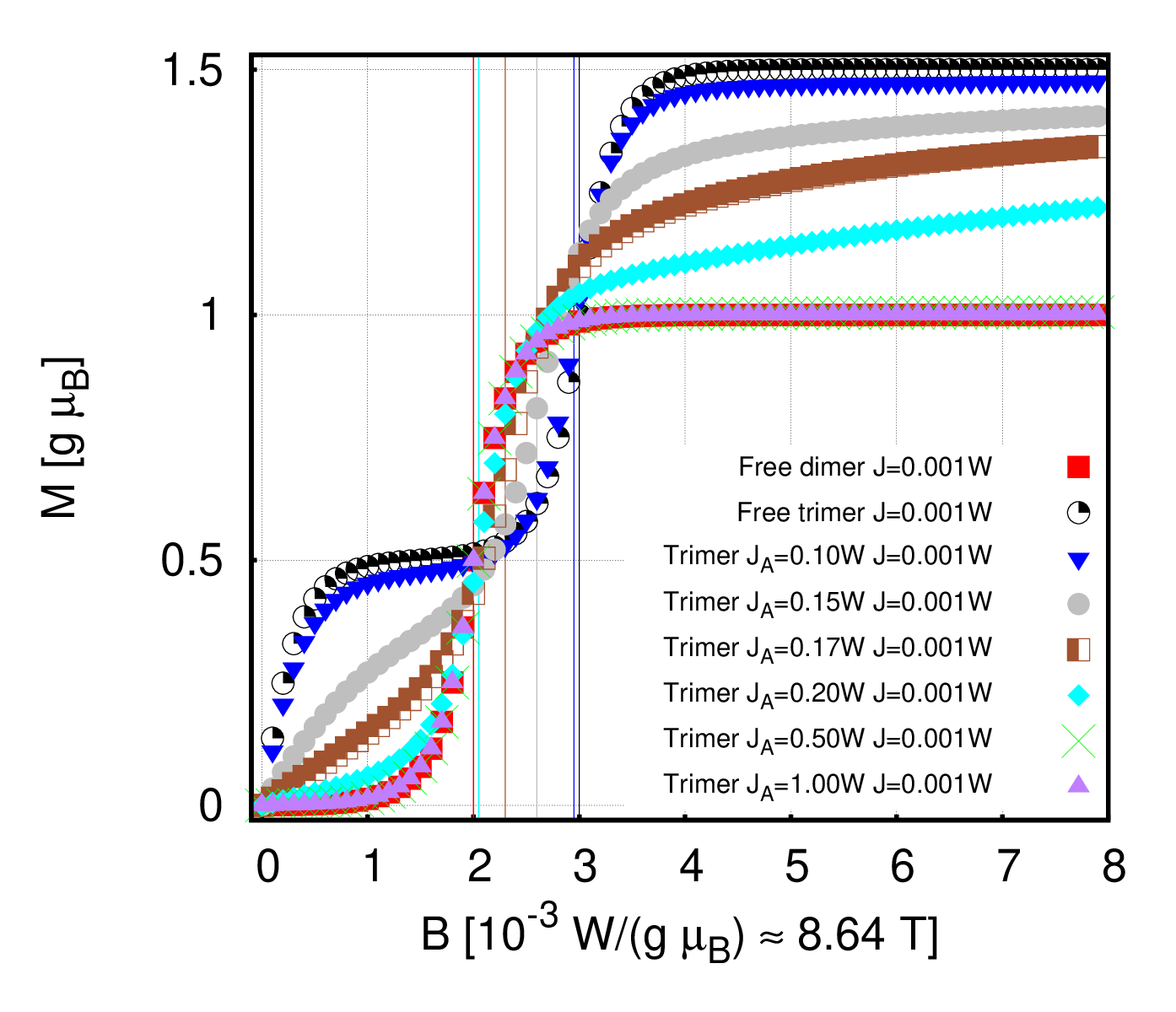}
\caption{(Color online) Impurity magnetization vs. magnetic
  field for the trimer shown in \figref{molsurf-f-a}~(a); 
  vertical lines mark the crossing fields for various
  scenarios; $T\approx 2 \cdot 10^{-4}\ W/k_B \approx 2.36$~K.}  
\label{trimer}
\end{figure}

After looking at the lowest Zeeman levels of a free trimer we
now consider the coupling of such a trimer to a surface as
described by Eq. (\ref{E-2-4}) and pictured in
\figref{molsurf-f-a}~(a). The strength of the coupling is
parametrized by $J_A$.  
The impurity magnetization curves depicted in Fig. \ref{trimer}
cover the whole range from the free ($J_A = 0$) or weakly
coupled case to the strongly coupled case, which is reached for
$J_A \gtrsim 0.5$~eV. While the case $J_A$ = 0 coincides with
the discussed free trimer by construction, the strongly coupled
case coincides with the magnetization curve of a free dimer. 

Additionally to the magnetization curves, Fig. \ref{trimer}
includes the crossing fields for various
scenarios which can be derived from the magnetization curves.  
The vertical lines marking these fields shift from the
analytical solution of a free trimer towards the analytical
solution of a free dimer for increasing coupling to the
substrate. 
The analytical solutions thus are boundaries for the crossing
field independent of the coupling $J_A$.  

\begin{figure}[ht!]
\centering
\includegraphics*[clip,width=85mm]{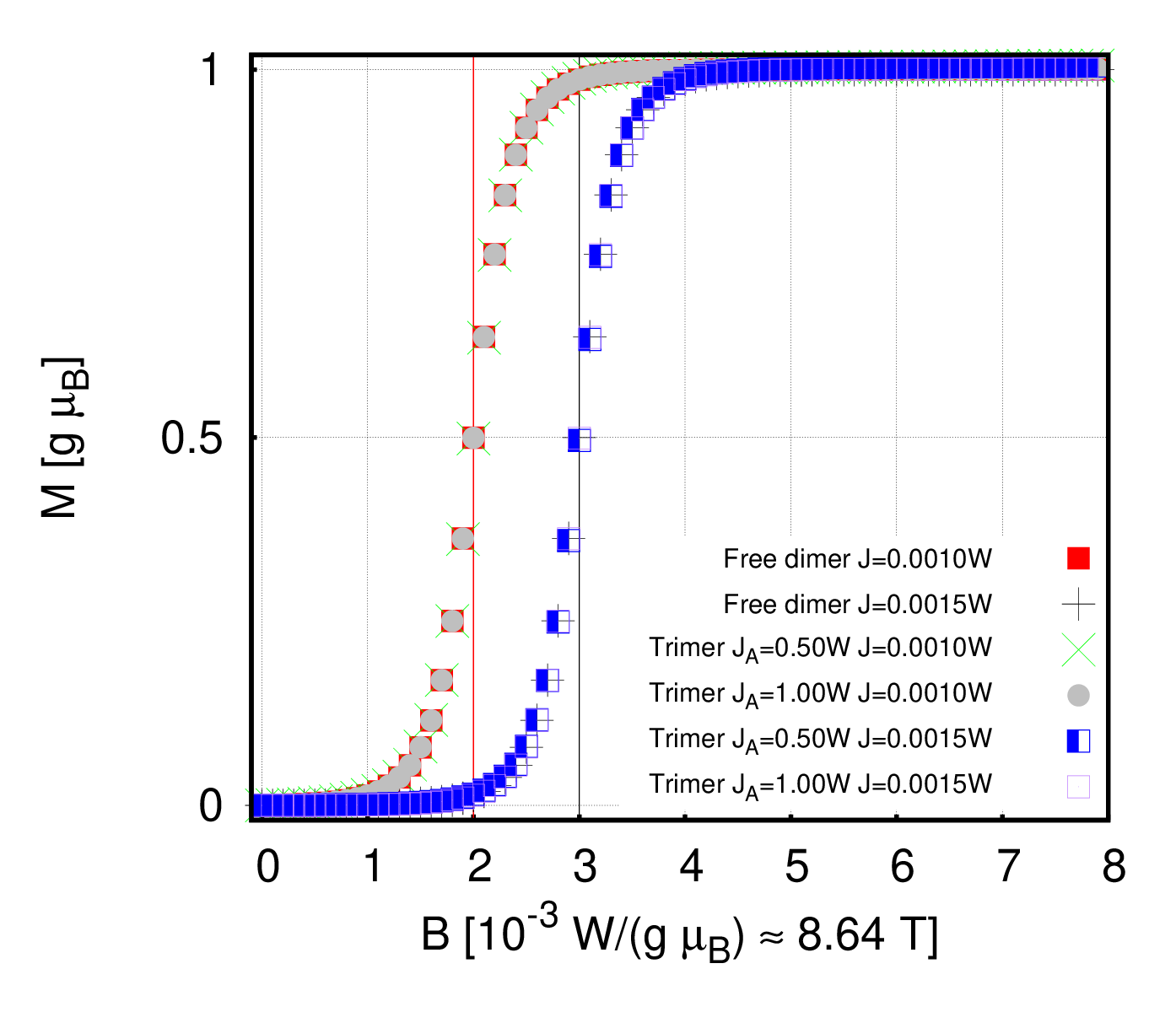}
\caption{(Color online) Impurity magnetization vs. magnetic
  field for the strong coupling case; vertical lines mark
  crossing fields; $T\approx 2 \cdot 10^{-4}\ W/k_B \approx
  2.36$~K. The magnetization curve depends only on $J$, not on
  $J_A$ in the strong coupling case.}  
\label{trimer15}
\end{figure}

Focusing on the strong coupling regime, \figref{trimer15} shows
variations of the couplings $J$ and $J_A$. 
Within this regime, i.e. for strong enough $J_A$, the
magnetization curves are independent of $J_A$ and coincide with
the solution for a free dimer and thus depend on the coupling
$J$ in the same way the analytical solution does. 
In particular, the crossing field is given by $B_c=2\cdot J/(g
\mu_B)$. Given a strong enough coupling to reach the strong
coupling regime it is therefore possible to derive $J$ from the
crossing field as done in Ref.~\onlinecite{CFJ:PRL08}.
Our investigation also shows that the maximum uncertainty in the
determination of $J$, in the case of unknown $J_A$, is given by the
difference between the (analytical) solutions for the crossing
field of the unscreened system and the fully screened one. 

\begin{figure}[ht!]
\centering
\includegraphics*[clip,width=85mm]{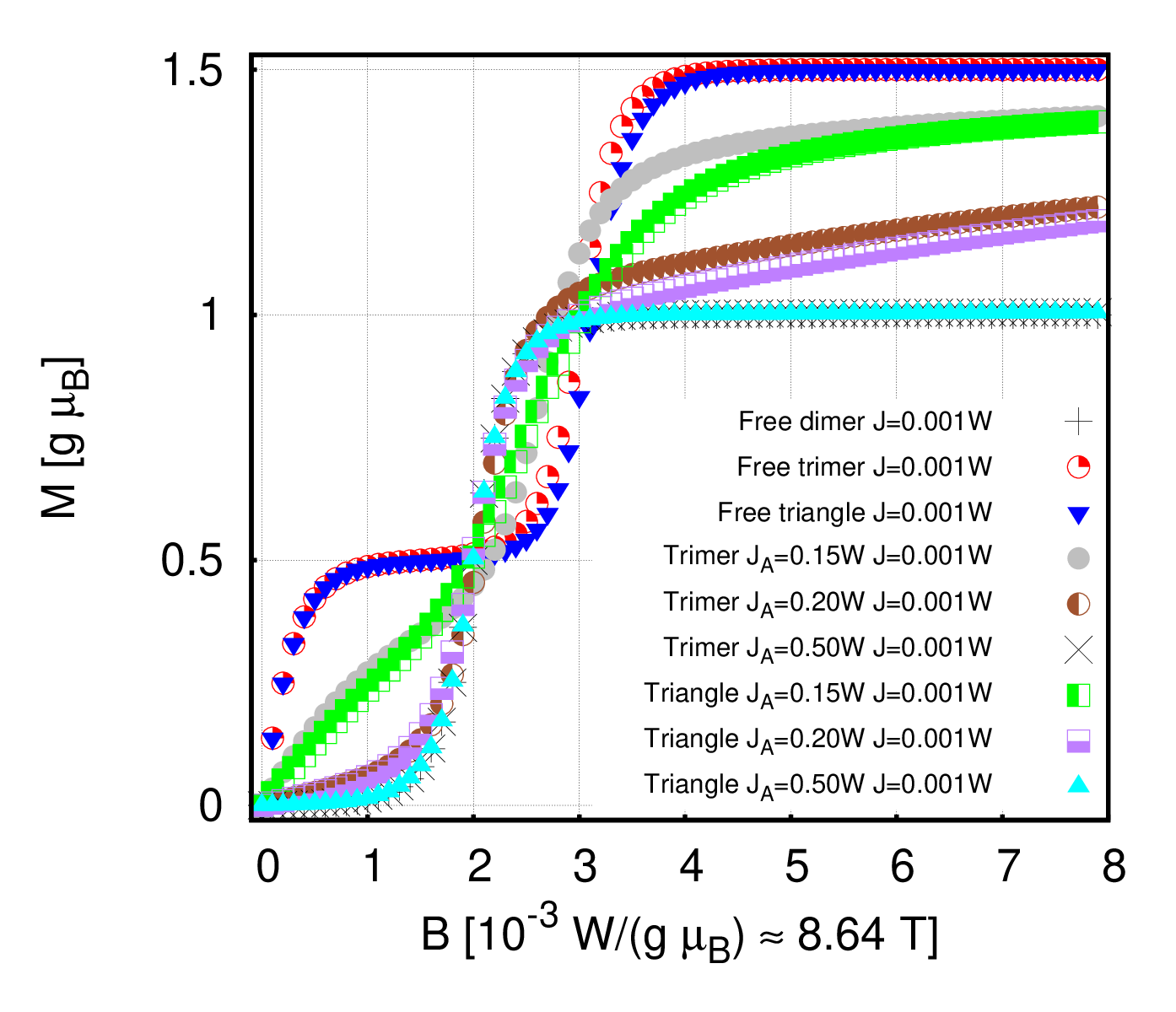}
\caption{(Color online) Impurity magnetization vs. magnetic
  field for trimeric and triangular impurities; $T \approx 2
  \cdot 10^{-4}\ W/k_B \approx 2.36$~K.}   
\label{trimertriangle}
\end{figure}

Similar results are obtained if a
triangular structure is used, where all three spins couple to
each other, but still only one couples to the substrate, compare
\figref{molsurf-f-a}~(b). In the
case of $J_A = 0$ analytical results can be obtained for both
trimer chain and triangle.\cite{MSL:PA00} These 
show that the magnetization curves and thus the crossing fields
coincide for T $\rightarrow$ 0 for both
systems. Figure~\ref{trimertriangle} shows the magnetization
curves of triangle and trimer chain for non-zero temperature where the
curves for $J_A = 0$ feature small differences. 
For intermediate couplings to the substrate, $0 < J_A <0.5$~eV,
the differences are more pronounced.

Figure \ref{trimertriangle} furthermore shows the case of strong
coupling to the substrate, again $J_A \gtrsim 0.5$~eV. In this
case trimer chain and triangle are indistinguishable on the
basis of their magnetization curves and thus their crossing
fields, compare $\times$-symbols and upright triangles in 
\figref{trimertriangle}.

\section{Relevant scales}
\label{sec-Ana}

To understand the transition to the strongly coupled case, as we have seen it in the results, we
take a look at the scales involved in the problem. The
Heisenberg spin Hamiltonian naturally features the internal
coupling $J$, which can be associated with an energy scale of

\begin{equation}
\Omega_{J} \approx  2 J \; .
\end{equation}

An additional energy scale dependent on the applied magnetic field arises from the Zeeman term

\begin{equation}
\Omega_B \approx  g \mu_B B \; .
\end{equation}

A third scale is tied to the temperature via

\begin{equation}
\Omega_{T} \approx k_B T \; .
\end{equation}

These energy scales now have to be compared to the energy scale
associated with the full screening, i.e. the energy scale of the
Kondo temperature

\begin{equation}
\Omega_{T_K} \approx  k_B T_K \; .
\end{equation}

The Kondo temperature itself can be estimated via its dependence
on the coupling to the conduction
band.\cite{Wil:RMP75,HoS:PRB13}  Within the context of our model
and unit system this means

\begin{equation}
\label{tkequation}
T_K = \sqrt{\frac{J_A}{W}} e^{-\frac{W}{J_A}} \frac{W}{k_B} \; ,
\end{equation}

so that the energy scale associated with it is given by

\begin{equation}
\Omega_{T_K} \approx  k_B T_K \approx \sqrt{J_A / W} \exp \left(\frac{-1}{J_A / W}\right) W \; .
\end{equation}

This means that varying the coupling strength effectively varies
the Kondo temperature in relation to the other energy scales.

The original situation, which we attempt to understand, can thus
be formulated with these scales. The observation of the crossing
field requires a variation of the magnetic field within the magnitude of
the internal interaction of the Heisenberg model. Furthermore
the temperature needs to be fixed sufficiently low in comparison
to both of these scales

\begin{equation}
\Omega_T < \Omega_{J} \lesssim \Omega_B \ll \Omega_{T_K} \; ,
\end{equation}

Thus the transition from the weakly to the strongly coupled case
can be described in terms of a variation of the scale
$\Omega_{T_K}$ in comparison to $\Omega_{J}$

\begin{eqnarray}
\Omega_{J} \ll \Omega_{T_K} \; ,
\end{eqnarray}

which translates to

\begin{eqnarray}
2 J_{I} \ll \sqrt{J_A / W} \exp \left(\frac{-1}{J_A / W}\right)  \; .
\end{eqnarray}

Given the general parameters of our model, the strength of the
internal interaction $J$ and the half-bandwidth W, we can thus derive a
criterion for the strongly coupled case

\begin{equation}
J_A \gg 0.2\ \text{eV} \; .
\end{equation}

This criterion should hold true regardless of the specific
structure of the spin cluster, as it is derived only from
general parameters. It is consistent with our observations for
the spin trimer and spin triangle as discussed above.

\begin{figure}[ht!]
\centering
\includegraphics*[clip,width=85mm]{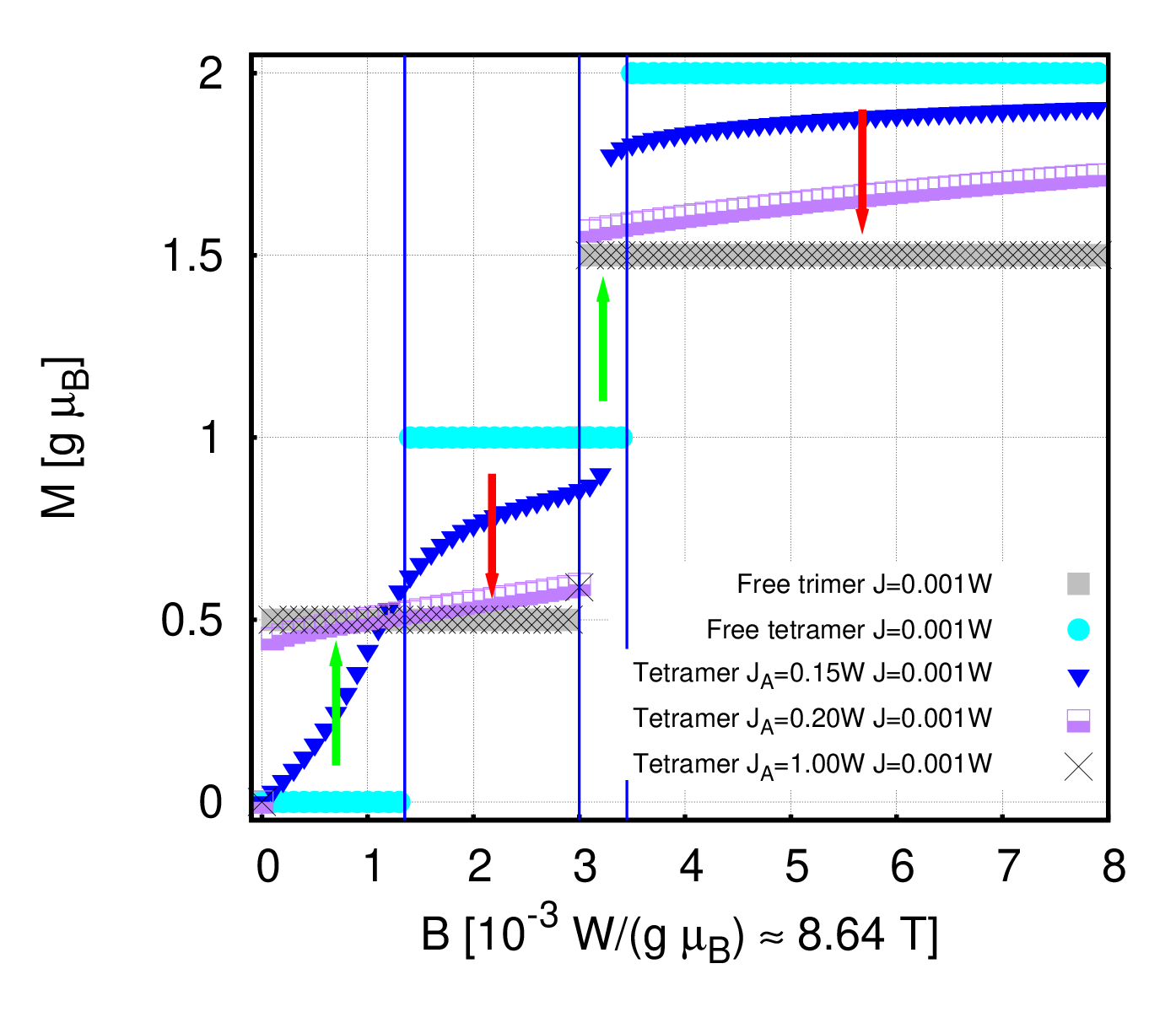}
\caption{(Color online) Impurity magnetization vs. magnetic
  field for tetramer impurities; vertical lines separate areas
  with an increase (green 
  arrows) or decrease (red arrows) in magnetization by strongly
  coupling the free system to the substrate; $T \approx 6 \cdot
  10^{-8}\ W/k_B \approx 6.92 \cdot 10^{-4}$~K.}  
\label{tetramer}
\end{figure}

\begin{figure}[ht!]
\centering
\includegraphics*[clip,width=85mm]{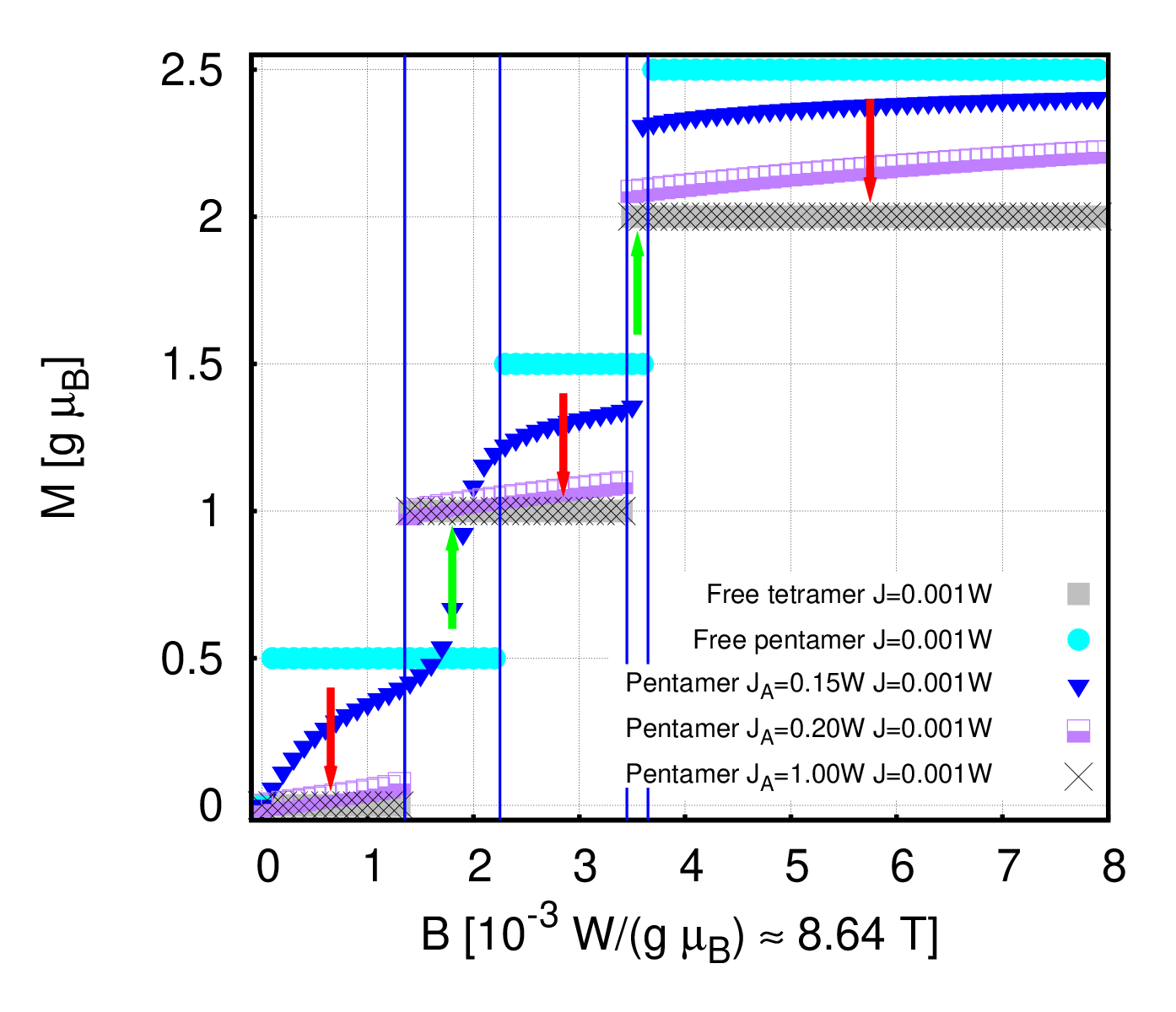}
\caption{(Color online) Impurity magnetization vs. magnetic
  field for pentamer impurities; vertical lines separate areas with an increase (green
  arrows) or decrease (red arrows) in magnetization by strongly
  coupling the free system to the substrate; $T \approx 6 \cdot
  10^{-8}\ W/k_B \approx 6.92 \cdot 10^{-4}$~K.}  
\label{pentamer}
\end{figure}

Finally, after our consideration of the scales in the problem,
we return to the chain-like impurities. 
We extend our original trimer chain to four and five spins as
also experimentally investigated in Ref.~\onlinecite{CFJ:PRL08}.
Figure \ref{tetramer} shows the magnetization curves of a
tetramer chain for various couplings to the substrate. 
As one might expect from the above considerations, the curves
are affected by the coupling to the substrate in a 
similar way to those of the trimer chain with the 
crossing fields decreasing for stronger couplings. 
In the strong coupling regime the tetramer chain then shows the
same magnetization curve and crossing field as the free
trimer.
Analogously the magnetization curve and crossing fields of the pentamer 
chain, as depicted in \figref{pentamer}, coincide with those of
the free tetramer in the strong coupling limit.

Unlike the other figures, Figures \ref{tetramer} and
\ref{pentamer} include vertical lines which confine field intervals of
(practically) constant magnetization for the 
free and the strongly coupled chains at a now very low
temperature of $T \approx 6 \cdot
  10^{-8}\ W/k_B \approx 6.92 \cdot 10^{-4}$~K. For those
  interval-ls an arrow 
indicates whether the magnetization is increased (green arrow up)
or decreased (red arrow down) due to the coupling to the substrate. For
the tetramer chain with its ground state spin of $S=0$ the
screening leads to an increased magnetization at (and thus a
response to) small magnetic fields. In contrast, the
magnetization for the pentamer chain with its ground state spin
of $S=1/2$ is decreased at small magnetic fields and thus will
not respond to small magnetic fields in the strong coupling
regime.  As would be expected for both the tetramer and the
pentamer chain after considering the scales of the problem as
done above, the strong coupling regime is only reached for $J_A
\gg 0.2$~eV.

\section{Summary}
\label{sec-4}

NRG calculations of the
impurity magnetization provide a very valuable tool in order to
rationalize experimental setups and results.
Future investigations of more
complicated impurities as for instance magnetic molecules with
non-Heisenberg terms in the Hamiltonian pose no problem. It is
also possible, although numerically demanding, to investigate
impurity problems with more than one channel, i.e. more than one
exchange contact to the surface by exploiting symmetry
properties.\cite{CBW:PRL09,ScS:PRB09,MSL:PRL12,ZMH:PRL12,HWC:PRB13,MGW:2013,Pavarini:205123}
For the structure of the problem at hand, we were able to see
how the crossing field, indicating level-crossings within the
energy spectrum, is influenced by a coupling of the spin cluster
to the conduction band. Including considerations of the involved
energy scales we come to the conclusion that within the model
used and given parameters for the internal interaction of scale
1~meV and half-bandwidth of 1eV, the strongly coupled case will
generally only occur for $J_A \gg 0.2$~eV.

\section*{Acknowledgment}

  This work was supported by the Deutsche Forschungsgemeinschaft
  through Research Unit 945. Computing time at the Leibniz
  Computing Center in Garching is gratefully acknowledged. We
  thank Martin H{\"o}ck for fruitful discussions and Theo Costi
  for drawing our attention to the successes of multi-channel
  calculations. 



%

\end{document}